\documentclass[preprint,12pt]{elsarticle}

\journal{Physica D}

\usepackage[T1]{fontenc}
\usepackage[latin2]{inputenc}
\usepackage{graphicx}

\begin{document}

\begin{frontmatter}

\title{Coherence and pattern formation in coupled logistic-map lattices}

\author[l1]{Maciej Janowicz}
\author[l2]{Arkadiusz Or{\l}owski}

\address[l1]{Institute of Physics of the Polish Academy of Sciences,
Aleja Lotnik\'ow 32/46, 02-668 Warsaw, Poland; e-mail: mjanow@ifpan.edu.pl}
\address[l2]{Faculty of Computer Science, Warsaw University of Life Sciences - SGGW,
ul. Nowoursynowska 159, 02-786 Warsaw, Poland}

\date{\today}

\begin{abstract}
Three quantitative measures of the spatiotemporal behavior of the coupled map lattices:
reduced density matrix, reduced wave function, and an analog of particle number, have been introduced. 
They provide a quantitative meaning to the concept of coherence which in the context of complex
systems have been used rather intuitively.
Their behavior suggests that the logistic coupled-map lattices approach the states
which resemble the condensed states of systems of Bose particles.
In addition, pattern formation in two-dimensional coupled map lattices based on the logistic mapping
has been investigated with respect to the non-linear parameter, the diffusion constant
and initial as well as boundary conditions.
\end{abstract}


\end{frontmatter}

\section{Introduction}
Coupled map lattices (CMLs) \cite{CF,Ilachinski} have long become a useful tool to investigate spatiotemporal
behavior of extended and possibly chaotic dynamical systems \cite{Kaneko1,WK,Kapral,Kaneko2}. It is so even though the most standard CML, that based on the coupling of logistic maps, is physically not particularly appealing as it is fairly remote
from any model of natural phenomena. Other, more complicated CMLs, have found interesting applications
in physical modeling. One should mention here CMLs developed to describe the Rayleigh-Benard
convection \cite{YK1}, dynamics of boiling \cite{Yanagita,GC}, formation and dynamics of clouds \cite{YK2},
crystal growth processes and hydrodynamics of two-dimensional flows \cite{Kaneko3}.

The most important characteristic quantities employed to study various types of CMLs include
co-moving Lyapunov spectra, mutual information flow, spatiotemporal power spectra, Kolmogorov-Sinai
entropy density, pattern entropy \cite{Kaneko3}. More recently, the detrended fluctuation analysis,
structure function analysis, local dimensions, embedding dimension and recurrence analysis have also
been introduced for CMLs \cite{MFMF}.

The purpose of this paper is to analyze the interesting features of the above-mentioned
most standard coupled map lattices which resemble the characteristisc of the condensates of Bose particles as well as those associated with formation of patterns in two spatial dimensions.
In particular, we investigate the formation of such patterns for relatively short times; their dependence
on two parameters which define CML as well as the initial conditions is found numerically.
Thus, the present work is very much in the spirit of classical papers \cite{Kapral,Kaneko2,Kaneko3}.
We believe, however, that the subject is very far from being exhausted as it is quite easy to find interesting patterns not 
discovered in the above works.  More importantly, we combine searching for interesting patterns with
the introduction of three additional quantities  with the help of which one can characterize the dynamics and statistical properties of CMLs. These are the reduced density matrix, the reduced wave function, and 
a quantity which is an analog of the number of particle. This is motivated, in part, by what we feel is the need 
to slightly deemphasize the connection of CMLs with finite-dimensional dynamical systems, and make 
their analysis similar to that of classical field theory, especially the Gross-Pitaevskii equation which is
used in the physics of Bose-Einstein condensation \cite{DGPS,Leggett}. 
Application of the classical field-theoretical methods in the physics of condensates have been described, 
e.g., in \cite{GGR1,GGR2,GGR3}.

Many interesting patterns emerge in the system while it still exhibits a transient behavior as can be seen,
e.g., in the plots of the ``number of particles". We have not attempted here to reach the regime
of stationary dynamics in each case. The problem for which times such a stationary regime becomes
established is beyond the scope of this work. We are content with the transient regime as long
as something interesting about the connection with condensates and about the patterns can be observed.
Let us notice that remarkable results on the transient behavior of extended systems
with chaotic behavior have been obtained, e.g., in \cite{Sinha,Janosi}. 

In addition, we observe that the condensate-like behavior has been reported in other systems
which are not connected with many Bose particles. Of particular interest are the
developments in the theory of complex networks \cite{BB,RB}. Here, however, we explore
the condensate-like behavior in the coupled map lattices.

The main body of this work is organized as follows. The mathematical model as well as the basic
definitions of reduced density matrix and reduced wave function are introduced in Section 2.
Section 3 provides a justification of our claim that the coupled map lattices based on logistic map 
exhibit properties which are analogous to those of the Bose-Einstein condensates (BEC).
The description of numerical results concerning pattern formation are contained in Section 4,
while Section 5 comprises a few concluding remarks.

\section{The model}

Let us consider the classical field $\psi(x, y, t)$ defined on a two-dimensional spatial lattice. Its evolution in (dimensionless, discrete) time $t$ is given by the following equation:

\begin{eqnarray}
\psi(x,y,t+1) &=& (1 - 4 d) f(\psi(x,y,t)) + d \left[ f(\psi(x+1,y,t)) + f(\psi(x-1,y,t)) \right. \nonumber \\
& + & \left. f(\psi(x,y+1,t)) \right.  + \left. f(\psi(x,y-1,t)) \right]
\end{eqnarray}

where the function $f$ is given by:

\begin{equation}
f(\psi) = c \psi (1 - \psi),
\end{equation}

and the parameters $c$ and $d$ are constant. The set of values taken by $\psi$ is the interval $[0, 1]$.
From the physical point of view the above diffusive model is rather bizarre, 
containing a field-dependent diffusion. 
There is no conserved quantity here which could play the role of energy or the number of excitations.

In the following the coefficient $d$ will be called the ``diffusion constant", and 
the coefficient $c$ - the ``non-linear parameter". 
It is assumed that $\psi$ satisfies either the periodic boundary or Dirichlet (with $\psi = 0$)
conditions on the borders of simulation box. The size of that box is $N \times N$. All our simulations have 
been performed with $N = 256$.

Let ${\tilde \psi}$ be the two-dimensional discrete Fourier transform of $\psi$,

\begin{equation}
{\tilde \psi}(m, n) = \sum_{x = 0}^{N-1} \sum_{y=0}^{N-1}
e^{2 \pi i m x/N} e^{2 \pi i n y/N} \psi(x, y), 
\end{equation}

Thus, ${\tilde \psi}$ may be interpreted as the momentum representation of the field $\psi$.

Below we investigate the relation between a CML described by Eq.(1) and a Bose-Einstein condensate.
Therefore, let us invoke the basic characteristics of the latter which are so important
that they actually form a part of its modern definition. 
These are \cite{PO,Yang}:

\begin{enumerate}
\item
The presence of off-diagonal long-range order (ODLRO).
\item
The presence of one eigenvalue of the one-particle reduced density matrix
which is much larger than all other eigenvalues.
\end{enumerate}

Let us notice that the property 2. corresponds to the well-known intuitive 
definition of the Bose-Einstein condensate. Namely, taking into account that 
the one-particle reduced density matrix $\rho^{(1)}$ has the following decomposition 
in terms of eigenvalues $\lambda_j$ and eigenvectors $|\phi_j \rangle$:

$$
\rho^{(1)} = \sum_j \lambda_j  |\phi_j \rangle \langle \phi_j|
$$

we realize that if one of the eigenvalues is much larger than the rest,
then the majority or at least a substantial fraction of particles 
is in the same single-particle quantum state.

In addition, for an idealized system of Bose particles with periodic boundary
conditions and without external potential, the following signature
of condensation is also to be noticed:

\begin{enumerate}
\setcounter{enumi}{2}
\item
The population of the zero-momentum mode is much larger than
population of all other modes.
\end{enumerate}

The properties 1. and 2. acquire quantitative meaning only if the 
one-particle reduced density matrix is defined. 
However, as our model is purely classical, the definition of 
that density matrix is not self-evident. Here
we make use of the classical-field approach to the theory of 
Bose-Einstein condensation \cite{GGR2,KGR} and define the quantities:
\begin{equation}
{\bar \rho}(x, x^{\prime}) = \langle \sum_{y = 0}^{N - 1} \psi(x, y) \psi(x^{\prime}, y) \rangle_{t}, 
\end{equation}

and 

\begin{equation}
\rho(x, x^{\prime}) = {\bar \rho}(x, x^{\prime}) / \sum_{x} {\bar \rho}(x, x).
\end{equation}

We shall call the quantity $\rho(x, x^{\prime})$ the reduced density matrix of CML.
The above definition in terms of an averaged quadratic form made of $\psi$ seems
quite natural, especially because $\rho$ is a real symmetric, positive-definite 
matrix with the trace equal to $1$. The sharp brackets $\langle ... \rangle_{t}$ denote
the time averaging:

$$
\langle (...) \rangle_{t} = \frac{1}{T_{s}} \sum_{t = T - T_{s}}^{T} (...),
$$

where $T$ is the total simulation time and $T_{s}$ is the averaging time.
In our numerical experiments $T$ has been equal to 3000, and $T_{s}$ has
been chosen to be equal to 300.

We can provide the quantitative meaning to the concept of off-diagonal 
long-range order (ODLRO) by saying that it is present in the system if 

$$
\rho(x_{1} + x, x_{1} - x)
$$

does not go to zero with increasing $x$ \cite{Yang}. If this is the case,
the system possesses the basic property 1. of Bose-Einstein condensates.

Let $W$ be the largest eigenvalue of $\rho$.
We will say that CML is in a "condensed state" if $W$ is significantly larger that all other
eigenvalues of $\rho$. If this is the case, the system possesses property 2.
of the Bose-Einstein condensates.
The corresponding eigenvector, $F(x)$, will be called
the reduced wave function of the (condensed part of) coupled map lattice.

One thing which still requires explanation is that the above definition of the 
reduced density matrix involves not only temporal, but also spatial averaging over $y$.
This is performed just for technical convenience, namely, to avoid dealing
with too large matrices. Strictly speaking, we are allowed to assess the presence
or absence of ODLRO only in one ($x$) direction. But that direction is arbitrary,
as we might equally well consider averaging over $x$ without any qualitative
change in the results. 

In the classical field theory the quantity ${\tilde \psi}^{\star} {\tilde \psi}$ represents
the particle density in momentum space; in the corresponding quantum theory,
upon the raising of $\psi$, $\psi^{\star}$ to the status of operators, 
${\tilde \psi}^{\star} {\tilde \psi}$ would be called the particle number operator.
Analogously, we introduce the number $P$ which - just for the purpose of the present
article - will be called the ``particle number", and is defined as:

\begin{equation}
P = \sum_{m = -N/2}^{N/2-1} \sum_{n=-N/2}^{N/2-1} |{\tilde \psi}(m,n)|^{2}.
\end{equation}

All the above definitions are modeled
after the corresponding definitions in the non-relativistic classical field
theory.

\section{Similarity to Bose-condensed systems}
 
We have performed our numerical experiment with six values of the non-linear parameter
$c$ ($3.5 + 0.1 \cdot i$, $i = 0, 1,...,5$), twenty five values of the diffusion constant
$d$ ($0.01 \cdot j$, $j = 1,2,..,25$), five different initial conditions, and two different
boundary conditions. The boundary conditions have been chosen either as periodic
or ``Dirichlet" ones, the latter with $\psi = 0$ at all boundaries. To save some space,
the tables below contain the results for $d$ being multiples of $0.05$, but the results
for other $d$ do not differ qualitatively from those reported below.
The following initial conditions have been investigated. The first - type (A) - initial
conditions are such that  $\psi(x, y, t)$ is ``excited" only at  a single point at $t = 0$:
$\psi(N/2, N/2, 0) = 0.5$, and $\psi(x, y, 0)$ is equal to zero at all other $(x,y)$.
Type (B) initial conditions are such that  $\psi(x, y, t)$ has initially
two non-vanishing values: $\psi(N/4, N/4, 0) = \psi(3 N/4, 3 N/4, 0) = 0.5$.
By type (C) initial conditions we mean those with $\psi(x, y, 0)$ being a Gaussian
function, $\psi(x, y, 0) = 0.5 \exp(-0.01 ((x - N/2)^2 + (y - N/2)^2))$. 
In type (D) initial conditions the Gaussian has been replaced with a sine function,
$\psi(x, y, 0) = 0.5 \sin(10 x/(N-1)$,
and type (E) are characterized by $\psi(x, y, 0)$ being equal to $0.5$ at all internal  points except
of one point - $(N/4, N/4)$ - where $\psi$ is equal to $0.50001$.

\subsection{Results for periodic boundary conditions}

Tables 1-5 show the dependence of the largest eigenvalue of the time-averaged reduced density
matrix on $c$ and $d$:

\begin{table}
\begin{center}
 \caption{Largest eigenvalue of the reduced density matrix. 
Periodic boundary conditions and type (A) initial conditions}
\vspace*{0.3cm}
 \begin{tabular}{|c|c|c|c|c|c|c|} \hline
 $ d \backslash c$ & 3.5 & 3.6 & 3.7 & 3.8 & 3.9 & 4.0\\
 \hline
 0.05 & 0.920 & 0.909 & 0.905 & 0.908 & 0.905 & 0.902 \\
 0.10 & 0.929 & 0.911 & 0.905 & 0.914 & 0.904 & 0.902\\
 0.15 & 0.948 & 0.917 & 0.929 & 0.945 & 0.907 & 0.904\\
 0.20 & 0.999 & 0.996 & 0.986 & 0.912 & 0.908 & 0.905\\
 0.25 & 0.499 & 0.496 & 0.493 & 0.483 & 0.454 & 0.453\\
 \hline
 \end{tabular}
\end{center}
\end{table}

\begin{table}
\begin{center}
 \caption{Largest eigenvalue of the reduced density matrix. 
Periodic boundary conditions and type (B) initial conditions}
\vspace*{0.3cm}
 \begin{tabular}{|c|c|c|c|c|c|c|} \hline
 $ d \backslash c$ & 3.5 & 3.6 & 3.7 & 3.8 & 3.9 & 4.0\\
  \hline
  0.05 & 0.921 & 0.909 & 0.906 & 0.909 & 0.905 & 0.902 \\
  0.10 & 0.940 & 0.918 & 0.902 & 0.923 & 0.904 & 0.902\\
  0.15 & 0.945 & 0.910 & 0.914 & 0.954 & 0.907 & 0.904\\
  0.20 & 0.912 & 0.911 & 0.898 & 0.899 & 0.908 & 0.905\\
  0.25 & 0.457 & 0.457 & 0.459 & 0.481 & 0.455 & 0.453\\
  \hline
 \end{tabular}
\end{center}
\end{table}

\begin{table}
\begin{center}
 \caption{Largest eigenvalue of the reduced density matrix. 
Periodic boundary conditions and type (C) initial conditions}
\vspace*{0.3cm}
 \begin{tabular}{|c|c|c|c|c|c|c|} \hline
 $ d \backslash c$ & 3.5 & 3.6 & 3.7 & 3.8 & 3.9 & 4.0\\
  \hline
  0.05 & 0.925 & 0.911 & 0.909 & 0.907 & 0.902 & 0.902 \\
  0.10 & 0.938 & 0.920 & 0.901 & 0.906 & 0.904 & 0.902\\
  0.15 & 0.953 & 0.928 & 0.910 & 0.915 & 0.906 & 0.905\\
  0.20 & 0.923 & 0.913 & 0.923 & 0.893 & 0.907 & 0.905\\
  0.25 & 0.910 & 0.903 & 0.888 & 0.909 & 0.906 & 0.904\\
  \hline
 \end{tabular}
\end{center}
\end{table}

\begin{table}
\begin{center}
 \caption{Largest eigenvalue of the reduced density matrix. 
Periodic boundary conditions and type (D) initial conditions}
\vspace*{0.3cm}
 \begin{tabular}{|c|c|c|c|c|c|c|} \hline
 $ d \backslash c$ & 3.5 & 3.6 & 3.7 & 3.8 & 3.9 & 4.0\\
  \hline
  0.05 & 0.927 & 0.913 & 0.901 & 0.899 & 0.893 & 0.858 \\
  0.10 & 0.934 & 0.924 & 0.911 & 0.904 & 0.896 & 0.887\\
  0.15 & 0.940 & 0.934 & 0.919 & 0.918 & 0.901 & 0.882\\
  0.20 & 0.940 & 0.932 & 0.925 & 0.924 & 0.902 & 0.881\\
  0.25 & 0.934 & 0.931 & 0.925 & 0.921 & 0.919 & 0.880\\
  \hline
 \end{tabular}
\end{center}
\end{table}

\begin{table}
\begin{center}
 \caption{Largest eigenvalue of the reduced density matrix. 
Periodic boundary conditions and type (E) initial conditions}
\vspace*{0.3cm}
 \begin{tabular}{|c|c|c|c|c|c|c|} \hline
 $ d \backslash c$ & 3.5 & 3.6 & 3.7 & 3.8 & 3.9 & 4.0\\
  \hline
  0.05 & 1.0 & 0.990 & 0.895 & 0.903 & 0.903 & 0.902 \\
  0.10 & 1.0 & 0.991 & 0.899 & 0.915 & 0.904 & 0.902\\
  0.15 & 1.0 & 0.990 & 0.914 & 0.906 & 0.907 & 0.904\\
  0.20 & 1.0 & 0.991 & 0.902 & 0.937 & 0.907 & 0.905\\
  0.25 & 1.0 & 0.994 & 0.931 & 0.918 & 0.903 & 0.453\\
  \hline
 \end{tabular}
\end{center}
\end{table}

There are several interesting observations which can be made in connection
with Tables 1-5. Firstly, with exception of the case $d = 0.25$ and arbitrary $c$ 
for type (A) initial conditions, the system exhibits one eigenvalue
of the reduced density matrix which is much larger than all other eigenvalues
for all other values of $c$ and $d$ and both types of initial conditions. 
This is one of the most important features of the Bose-condensed
matter, as explained in Section 2. Our system clearly has the property 2. of BEC.
Secondly, for the case $d = 0.25$ and types (A) and (B) initial conditions, 
the largest eigenvalue is slightly lower than $0.5$. We have checked that, for each $c$, there are 
{\em two} almost equal eigenvalues which are much larger than all other eigenvalues.
The presence of two such eigenvalues of the reduced density matrix 
also has its analog in the physics of Bose-Einstein condensation; it is characteristic
for the so-called quasi-condensates  \cite{PSW1,PSW2,KGR}. Further,
it seems there are certain regularities in the $c$ and $d$ dependence of the maximal
eigenvalue. In most (but not all) cases, the value of $W$ appears
to decrease with growing $c$  for given $d$. In all cases except of $d=0.25$,  
$W$ has had the largest value for $c$ equal to 3.5, that is, below the threshold of chaos 
for a single logistic map.
The fact that the largest eigenvalue for type (E) initial conditions does not practically
differ from $1$ for $c = 3.5$ and is still very close to $1$ for $c = 3.6$ can obviously
be attributed to the fact that those values of the nonlinear parameters ar too "weak"
to introduce sufficient variation into the system which is initially almost perfectly
homogeneous. 

In Figure 1(a-b) we have displayed the spatial dependence of the quantity ("one-particle
correlation function") $\sigma(x) = \rho(N/2 + x, N/2 - x)$ for $x = 0,1,2,...,N/2-1$,
$d = 0.20$, $c = 3.7$, periodic boundary conditions, and five types
of initial conditions.

\begin{figure}[h!]
\begin{center}
\includegraphics[angle=0,width=11.0cm,height=10.0cm]{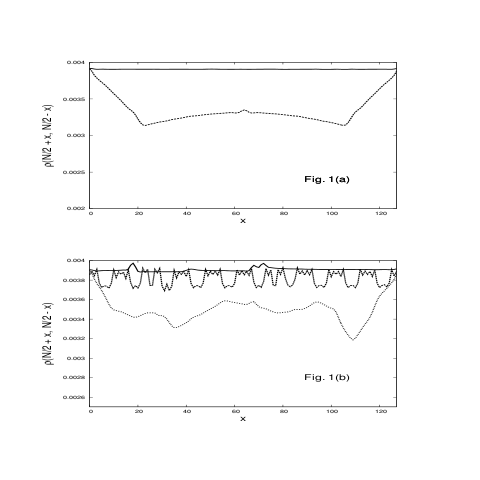}
\caption{Spatial dependence of the one-particle correlation functions for 
$d = 0.20$, $c = 3.7$ and periodic boundary conditions; 
(a) solid line:  type (A) initial conditions, dashed line: type (B) boundary conditions;
(b) solid line:  type (C) initial conditions, dashed line: type (D) boundary conditions, 
dotted line: type (E) initial
conditions.}
\end{center}
\end{figure}

While the values of the above "one-particle correlation function" for $x = 0$ and $x = N/2$
must be equal due to the boundary conditions, a strong decrease of $\sigma(x)$ 
for $x$ being far from $0$ or $N/2$ would have to take place if there were {\em no} long-range order.
However, $\sigma(x)$ never falls below the $75\%$ of its value for $x = 0$.
In addition, for types (A), (C) and (D) initial conditions the change of $\sigma$ with $x$
reduces itself to very small fluctuations. 
We can conclude that our system exhibits the property 1. of the Bose-Einstein condensates.
Still, the considerable variation of $\sigma(x)$ wwith respect to initial conditions
seems quite interesting and probably deserves some further investigation.

The plots in Fig. 2(a-b) illustrate the dependence of eigenvectors $F(x)$ 
("wave functions of the condensate") corresponding to the largest eigenvalue $W$  
on $x$  for all types of initial conditions. The most important feature of those 
plots is very weak dependence of $F(x)$ on $x$, with a single exception of type (D)
initial conditions where the variation slightly exceeds $10\%$.
This feature has also been observed for all other values of our parameters
and for much longer times as well, regardless of the final state of the system.

\begin{figure}[h!]
\begin{center}
\includegraphics[angle=0,width=11.0cm,height=10cm]{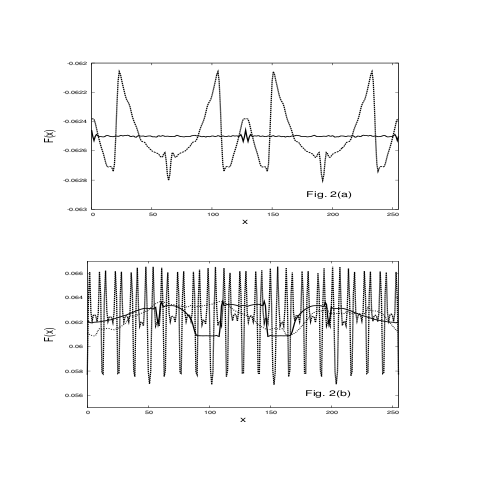}
\caption{Spatial dependence of the eigenvector, corresponding to the largest eigenvalue
of the reduced density matrix for $d = 0.20$ and $c = 3.7$ and periodic boundary conditions;  
(a) solid line:  type (A) initial conditions, dashed line: type (B) initial conditions
(b) solid line:  type (C) initial conditions, 
dashed line: type (D) initial conditions, dotted line: type (E) initial conditions.}
\end{center}
\end{figure}

Thus, one may say that the correlation length is virtually infinite,
which, again, is a characteristic feature of the strongly condensed physical systems.
We note that this is true even for the $d = 0.25$ and type (A) initial conditions, where
the system resembles quasi-condensates. In such a case the density fluctuations 
should not differ from those of the ``true" condensates; the difference lies in the phase
fluctuations. Elaboration of that interesting point is, however, beyond the scope
of the present work.

To make our case of pointing out the CML resemblance to Bose condensates even stronger,
we have checked the behavior of the field $\psi$ in momentum space.
In Figures 3(a-e) the plots of the moduli $|{\tilde \psi}|$ as functions
of two components of their ``momentum'' argument are shown for periodic boundary conditions
and five types of initial conditions. The function $|{\tilde \psi}(m, n)|$ is normalized in such a way 
that its maximal value is $1$.

\begin{figure}[h!]
\begin{center}
\includegraphics[angle=0,width=11.0cm,height=14.0cm]{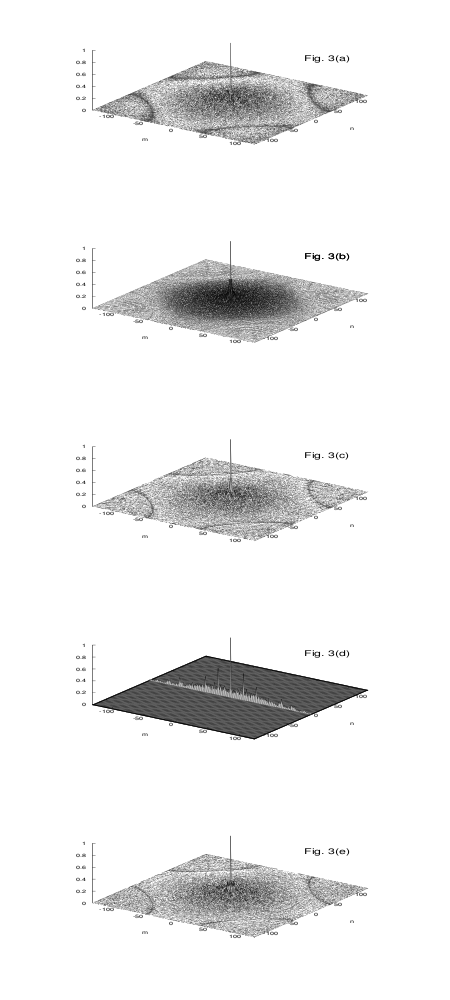}
\caption{The dependence of $|{\tilde \psi}|$ on the discrete vector of momentum $(m,n)$
for $d = 0.20$, $c = 3.7$, and periodic boundary conditions. The values of
 $|{\tilde \psi}|$ has been normalized in such a way that $|{\tilde \psi}(0,0)| = 1$;
(a) type (A) initial conditions; (b) type (B) initial condition; (c) type (C) initial
conditions; (d) type (D) initial conditions; (e) type (E) initial conditions.}
\end{center}
\end{figure}

Almost all plots in Fig. 3 are qualitatively the same except, again, of that corresponding to the 
sinusoidal initial conditions. In addition, they are representative for 
the entire spectrum of values of $c$ and $d$,
even for $d = 0.25$ with type (A) initial conditions (that is, for ``quasi-condensates'').
Strong peak at the zero momentum clearly dominates all the other maxima. The fact
that the zeroth mode is the only one which is so strongly populated is yet another
feature of Bose-condensed system of particles - our system exhibits the property 3.
of condensates. However, the lateral amplitudes for the sinusoidal initial conditions
are relatively high, as can be seen in Fig. 3(d). Although even for this case the 
dominance of the central mode is clear, it appears that it is weaker than in the case
of other initial conditions.
 
Let us finally consider the function $P(t)$, which is an analog of the particle number.
In our system $P(t)$ is a genuine function of time, and there is no conservation law for it.
 
The figure 4(a-e) contains several plots of time dependence of $P(t)$ for $d = 0.020$ and $c = 3.7$,
periodic boundary conditions and five types of initial conditions.

\begin{figure}[h!]
\begin{center}
\includegraphics[angle=0,width=11.0cm,height=15.0cm]{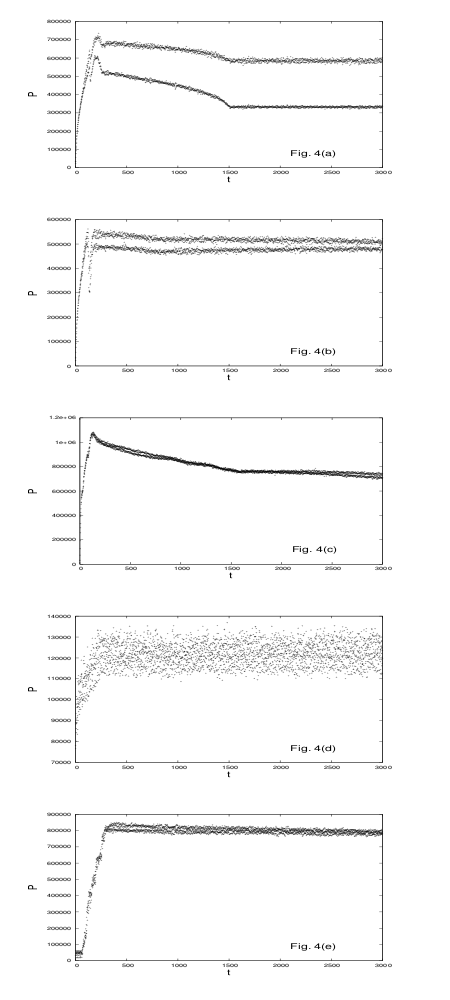}
\caption{Time dependence of the particle number $P$ for $d = 0.20$,  $c = 3.7$, 
and periodic boundary conditions; (a) type (A) initial conditions,
(b) type (B) initial conditions, (c) type (C) initial conditions,
(d) type (D) initial conditions, (e) type (E) initial conditions.} 
\end{center}
\end{figure}

Let us first observe that the asymptotic dynamics of $P(t)$ for large $t$ which are very different
for does indeed depend on initial conditions. The ``bands" which are very characteristic
for type (A) and (B) initial conditions practically disappear for type (C) and (E) initial
conditions. The sinusoidal (type (D)) initial conditions are again quite special for two reasons.
Not only is the change of $P(t)$ very erractic with no visible ``bands", but its mean value is,
in addition, almost one order of magnitude smaller than in the case of other initial
conditions. So far, we cannot offer any explanation of this feature.

\subsection{Results for Dirichlet boundary conditions} 

Tables 6-10 show the dependence of the largest eigenvalue of the time-averaged reduced density
matrix on $c$ and $d$:

\begin{table}
\begin{center}
 \caption{Largest eigenvalue of the reduced density matrix. 
Dirichlet boundary conidtions and type (A) initial conditions}
\vspace*{0.3cm}
 \begin{tabular}{|c|c|c|c|c|c|c|} \hline
 $ d \backslash c$ & 3.5 & 3.6 & 3.7 & 3.8 & 3.9 & 4.0\\
 \hline
 0.05 & 0.920 & 0.909 & 0.904 & 0.908 & 0.905 & 0.902 \\
 0.10 & 0.922 & 0.916 & 0.913 & 0.911 & 0.903 & 0.901\\
 0.15 & 0.940 & 0.944 & 0.929 & 0.933 & 0.905 & 0.903\\
 0.20 & 0.997 & 0.997 & 0.986 & 0.909 & 0.906 & 0.905\\
 0.25 & 0.499 & 497 & 0.493 & 0.483 & 0.453 & 0.453\\
 \hline
 \end{tabular}
\end{center}
\end{table}

\begin{table}
\begin{center}
 \caption{Largest eigenvalue of the reduced density matrix. 
Dirichlet boundary conidtions and type (B) initial conditions}
\vspace*{0.3cm}
 \begin{tabular}{|c|c|c|c|c|c|c|} \hline
 $ d \backslash c$ & 3.5 & 3.6 & 3.7 & 3.8 & 3.9 & 4.0\\
  \hline
  0.05 & 0.920 & 0.909 & 0.906 & 0.910 & 0.905 & 0.902 \\
  0.10 & 0.944 & 0.915 & 0.903 & 0.910 & 0.904 & 0.901\\
  0.15 & 0.950 & 0.973 & 0.899 & 0.920 & 0.905 & 0.904\\
  0.20 & 0.953 & 0.956 & 0.958 & 0.916 & 0.906 & 0.905\\
  0.25 & 0.486 & 0.489 & 0.483 & 0.483 & 0.453 & 0.453\\
  \hline
 \end{tabular}
\end{center}
\end{table}

\begin{table}
\begin{center}
 \caption{Largest eigenvalue of the reduced density matrix. 
Dirichlet boundary conidtions and type (C) initial conditions}
\vspace*{0.3cm}
 \begin{tabular}{|c|c|c|c|c|c|c|} \hline
 $ d \backslash c$ & 3.5 & 3.6 & 3.7 & 3.8 & 3.9 & 4.0\\
  \hline
  0.05 & 0.925 & 0.912 & 0.909 & 0.907 & 0.902 & 0.902 \\
  0.10 & 0.937 & 0.920 & 0.902 & 0.897 & 0.904 & 0.901\\
  0.15 & 0.948 & 0.928 & 0.907 & 0.903 & 0.905 & 0.904\\
  0.20 & 0.923 & 0.913 & 0.923 & 0.906 & 0.906 & 0.905\\
  0.25 & 0.911 & 0.904 & 0.887 & 0.894 & 0.906 & 0.904\\
  \hline
 \end{tabular}
\end{center}
\end{table}

\begin{table}
\begin{center}
 \caption{Largest eigenvalue of the reduced density matrix. 
Dirichlet boundary conidtions and type (D) initial conditions}
\vspace*{0.3cm}
 \begin{tabular}{|c|c|c|c|c|c|c|} \hline
 $ d \backslash c$ & 3.5 & 3.6 & 3.7 & 3.8 & 3.9 & 4.0\\
  \hline
  0.05 & 0.913 & 0.912 & 0.900 & 0.898 & 0.896 & 0.902 \\
  0.10 & 0.936 & 0.926 & 0.916 & 0.897 & 0.904 & 0.901\\
  0.15 & 0.942 & 0.936 & 0.925 & 0.911 & 0.905 & 0.904\\
  0.20 & 0.942 & 0.934 & 0.929 & 0.895 & 0.906 & 0.905\\
  0.25 & 0.934 & 0.931 & 0.964 & 0.936 & 0.905 & 0.904\\
  \hline
 \end{tabular}
\end{center}
\end{table}

\begin{table}
\begin{center}
 \caption{Largest eigenvalue of the reduced density matrix. 
Dirichlet boundary conidtions and type (E) initial conditions}
\vspace*{0.3cm}
 \begin{tabular}{|c|c|c|c|c|c|c|} \hline
 $ d \backslash c$ & 3.5 & 3.6 & 3.7 & 3.8 & 3.9 & 4.0\\
  \hline
  0.05 & 1.0 & 0.993 & 0.922 & 0.926 & 0.913 & 0.902 \\
  0.10 & 1.0 & 0.992 & 0.922 & 0.900 & 0.903 & 0.901\\
  0.15 & 1.0 & 0.992 & 0.929 & 0.906 & 0.905 & 0.904\\
  0.20 & 1.0 & 0.992 & 0.922 & 0.913 & 0.906 & 0.905\\
  0.25 & 1.0 & 0.991 & 0.922 & 0.904 & 0.906 & 0.904\\
  \hline
 \end{tabular}
\end{center}
\end{table}

The most important conclusion one can draw from the Tables (1-5) is the 
same as in the case of periodic boundary conditions: there exist one 
dominant eigenvalues for almost values parameters except of the value $d=0.25$
where exactly two dominant eigenvalues are present. This feature suggests
that the change of boundary conditions does not diminish the similarity
of our system to the Bose-Einstein condensates, or quasi-condensates.
Also, it seems that the general trend of change (namely, decrease) of the 
largest eigenvalue with growing $c$ for given $d$ is here observed, but again,
the number of exceptions is considerable.

In Figure 5 we have displayed the spatial dependence of the quantity ("one-particle
correlation function") $\sigma(x) = \rho(N/2 + x, N/2 - x)$ for $x = 0,1,2,...,N/2-1$,
$d = 0.20$, $c = 3.7$, periodic boundary conditions, and five types
of initial conditions. As in the case of periodic boundary conditions, the long-range
order is very transparent. The function $\rho(N/2 - x, N/2 + x)$ approaches zero only
for the values of its arguments approaching boundaries.

\begin{figure}[h!]
\begin{center}
\includegraphics[angle=0,width=11.0cm,height=13.0cm]{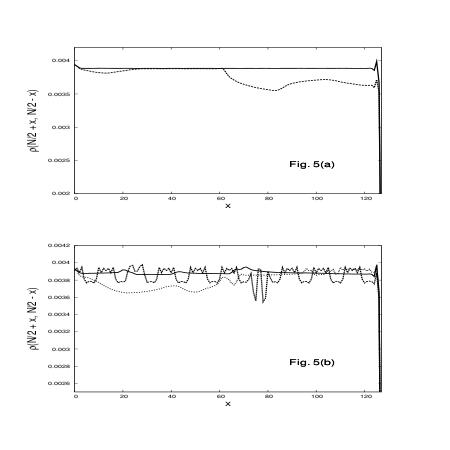}
\caption{Spatial dependence of the one-particle correlation functions for 
$d = 0.20$, $c = 3.7$ and Dirichlet boundary conditions; 
(a) solid line:  type (A) initial conditions, dashed line: type (B) initial conditions;
(b) solid line:  type (C) initial conditions, dashed line: type (D) boundary conditions, 
dotted line: type (E) initial conditions.}
\end{center}
\end{figure}

Figure 6(a-c) contains the plots of the eigenvectors ("wave functions of the condensate") corresponding
to the largest eigenvalue $W$ for all types of initial conditions.

\begin{figure}[h!]
\begin{center}
\includegraphics[angle=0,width=11.0cm,height=14cm]{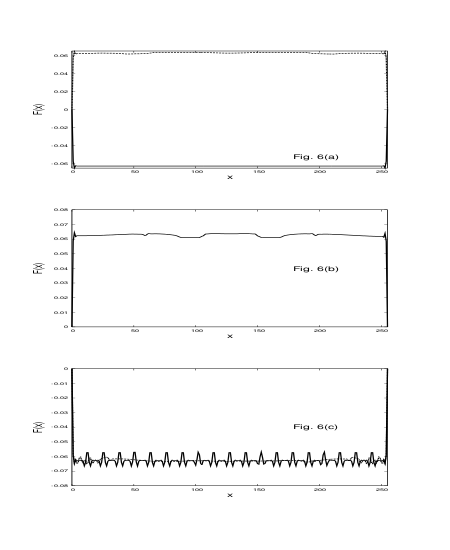}
\caption{Spatial dependence of the eigenvector, corresponding to the largest eigenvalue
of the reduced density matrix for $d = 0.20$ and $c = 3.7$ and Dirichlet boundary conditions;  
(a) solid line:  type (A) initial conditions, dashed line: type (B) initial conditions
(b) solid line:  type (C) initial conditions, 
dashed line: type (D) initial conditions, dotted line: type (E) initial conditions.}
\end{center}
\end{figure}

All the above plots are very flat, except of the arguments close to the boundaries,
and the values of the "wave functions" never never approach zero. The system appears
to be globally correlated, even though it is still in the transient regime with only
local synchronization.

As in the previous Section, we have checked the behavior of the field $\psi$ in momentum space.
In Figures 7(a-e) the plots of the moduli $|{\tilde \psi}|$ as functions
of two components of their ``momentum'' argument are shown for two types of initial
conditions. The function $|{\tilde \psi}(m, n)|$ is normalized in such a way that its maximal
value is $1$.

\begin{figure}[h!]
\begin{center}
\includegraphics[angle=0,width=11.0cm,height=15.0cm]{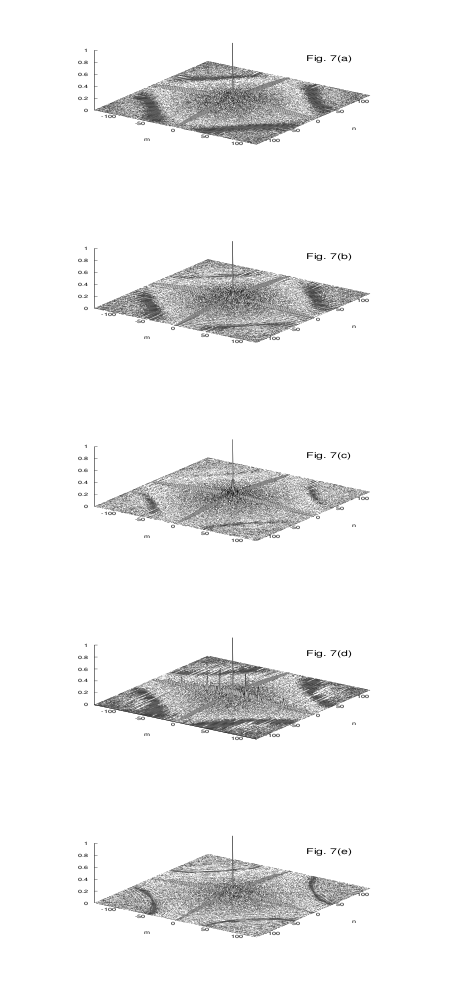}
\caption{The dependence of $|{\tilde \psi}|$ on the discrete vector of momentum $(m,n)$
for $d = 0.20$, $c = 3.7$, and Dirichlet boundary conditions. The values of
 $|{\tilde \psi}|$ has been normalized in such a way that $|{\tilde \psi}(0,0)| = 1$;
(a) type (A) initial conditions; (b) type (B) initial condition; (c) type (C) initial
conditions; (d) type (D) initial conditions; (e) type (E) initial conditions.}
\end{center}
\end{figure}

The dominance of a single $(m = 0, n = 0)$ mode is transparent,
except that in the case the type (D) (sinusoidal) initial conditions the population
of lateral modes is substantially bigger that in the other cases.

Figure 8(a-e) contains the plots of time dependence of the variable $P$ for $d = 0.020$ and $c = 3.7$.

\begin{figure}[h!]
\begin{center}
\includegraphics[angle=0,width=11.0cm,height=13.0cm]{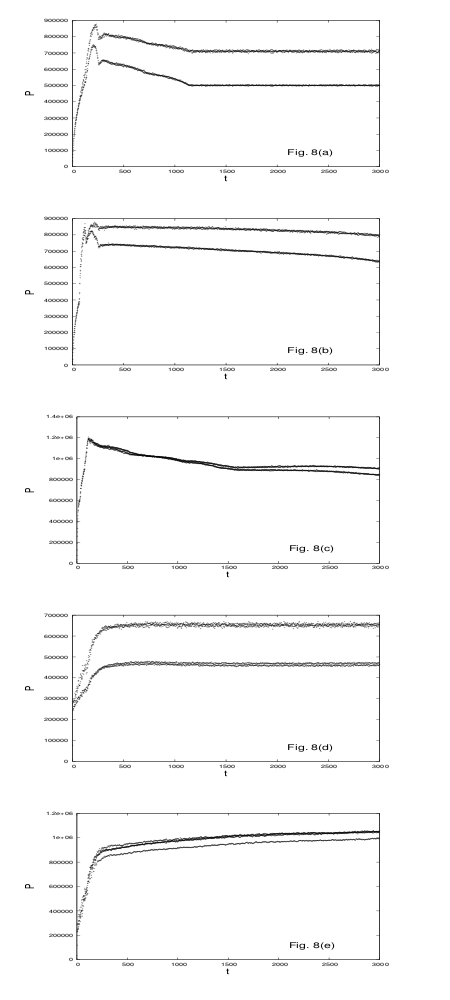}
\caption{Time dependence of the particle number $P$ for $d = 0.20$,  $c = 3.7$, 
and Dirichlet boundary conditions; (a) type (A) initial conditions,
(b) type (B) initial conditions, (c) type (C) initial conditions,
(d) type (D) initial conditions, (e) type (E) initial conditions.} 
\end{center}
\end{figure}

It seems that the dynamics of the particle is quite sensitive to the boundary conditions
for we can see here substantial deviation from the dynamics of $P$ in the case
of periodic boundary conditions. In particular, in the time regime which is 
investigated here, the ``bands" appear to be far better visible for the Dirichlet
boundary conditions. What is more, the case of sinusoidal initial conditions is no longer
much different from all the others, although the mean number of particles
is still smallest in that case.

\section{Large-scale pattern formation}

We have observed the following general rules in the process of pattern formation
in our system. Firstly, the patterns are incomparably better developed (much better
visible) for any ``structured" initial conditions (like those considered in this work)
than in the case of random initial conditions. The initial inhomogeneities (or ``seeds")  
serve the building of large structures much better than fully random conditions, 
which is fairly intuitive. The patterns are best developed for smaller values of the non-linear parameter
and intermediate values of the diffusion constant.

In Figs. 9-15 we show shaded-contour plots representing the values of the field
$\psi(x,y)$ after 3000 time steps for periodic boundary conditions and types (A-C) and (E) initial conditions. 
There are no figures for type (D) (sinusoidal) boundary conditions because they are quite
uninteresting, displaying merely the stripes corresponding to the sinusoidal initial ``excitation".

\begin{figure}
\includegraphics[width=4.5cm,height=4.5cm]{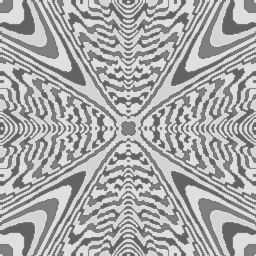}
\includegraphics[width=4.5cm,height=4.5cm]{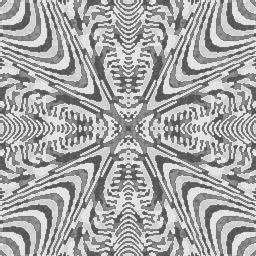}
\includegraphics[width=4.5cm,height=4.5cm]{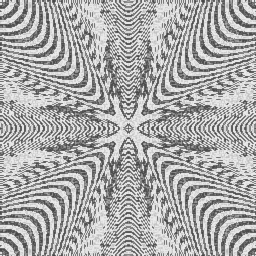}
\caption{Grayscale shaded contour graphics representing the values of the field $\psi$ after
3000 time steps for $d = 0.05$, periodic boundary conditions, and three values of $c$ for type (A) initial conditions;
(a)$c = 3.5$, (b)$c = 3.6$, (c)$c = 3.7$.
Brighter regions are those with higher values of $\psi$. }
\end{figure}

Naturally, the large structures visible in Figs. 9-14 reflect, to some extent, 
the symmetry of the simulation box. More interesting observation is that the change 
from periodic ($c = 3.5$) to chaotic ($c = 3.6$) regime - 
as defined for individual maps -  does not lead,  
in the case of very slow diffusion, to any spectacular change of the pattern.

\begin{figure}[h!]
\includegraphics[width=4.5cm,height=4.5cm]{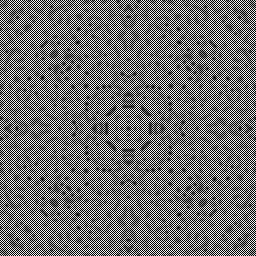}
\includegraphics[width=4.5cm,height=4.5cm]{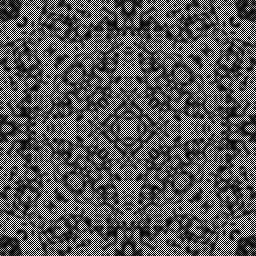}
\includegraphics[width=4.5cm,height=4.5cm]{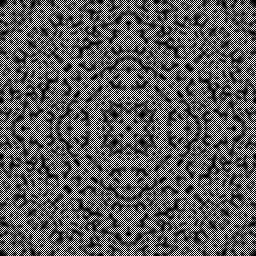}
\caption{Grayscale shaded contour graphics representing the values of the field $\psi$ after
3000 time steps for $d = 0.25$, periodic boundary conditions and 
three values of $c$ and type (A) initial conditions; (a) $c = 3.8$, (b) $c = 3.9$, (c) $c = 4.0$.
Brighter regions are those with higher values of $\psi$. }
\end{figure}

\begin{figure}[h!]
\includegraphics[width=4.5cm,height=4.5cm]{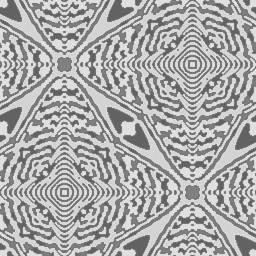}
\includegraphics[width=4.5cm,height=4.5cm]{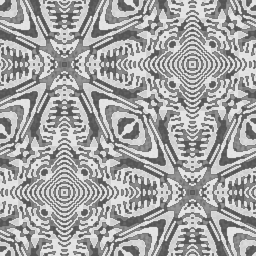}
\includegraphics[width=4.5cm,height=4.5cm]{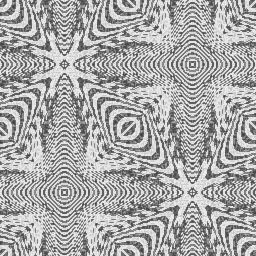}
\caption{Grayscale shaded contour graphics representing the values of the field $\psi$ after
3000 time steps for $d = 0.05$, periodic boundary conditions, 
three values of $c$, type (B) initial conditions; (a) c = 3.5, (b) c = 3.6, (c) c = 3.7.
Brighter regions are those with higher values of $\psi$. }
\end{figure}

\begin{figure}[h!]
\includegraphics[width=4.5cm,height=4.5cm]{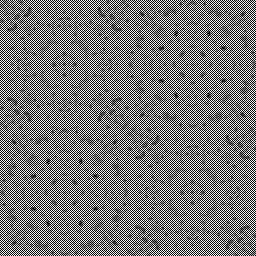}
\includegraphics[width=4.5cm,height=4.5cm]{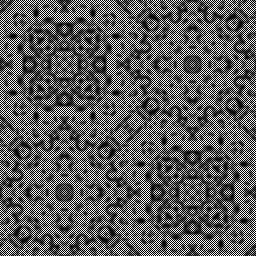}
\includegraphics[width=4.5cm,height=4.5cm]{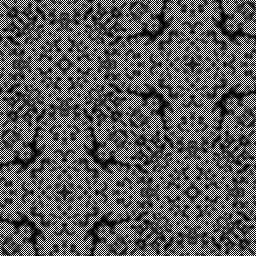}
\caption{Grayscale shaded contour graphics representing the values of the field $\psi$ after
3000 time steps for $d = 0.25$, periodic boundary conditions, three values of $c$ and type (B) initial conditions;
(a) $c = 3.8$, (b) $c = 3.9$, (c) $c = 4.0$.
Brighter regions are those with higher values of $\psi$. }
\end{figure}

\begin{figure}[h!]
\includegraphics[width=4.5cm,height=4.5cm]{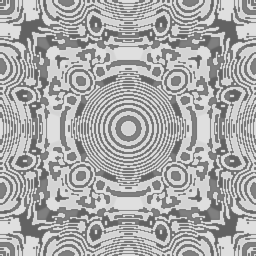}
\includegraphics[width=4.5cm,height=4.5cm]{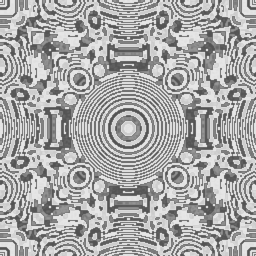}
\includegraphics[width=4.5cm,height=4.5cm]{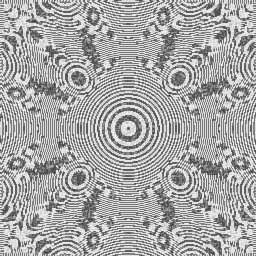}
\caption{Grayscale shaded contour graphics representing the values of the field $\psi$ after
3000 time steps for $d = 0.05$, periodic boundary conditions, 
three values of $c$, type (C) initial conditions; (a) c = 3.5, (b) c = 3.6, (c) c = 3.7.
Brighter regions are those with higher values of $\psi$.}
\end{figure}

The most characteristic feature of the fast-diffusion (i.e. large $d$) case is the disappearance of
the large-scale structures even for any type of initial conditions. 
However, somewhat more pronounced grainy structures reappear for $c = 3.9$.

\begin{figure}[h!]
\includegraphics[width=4.5cm,height=4.5cm]{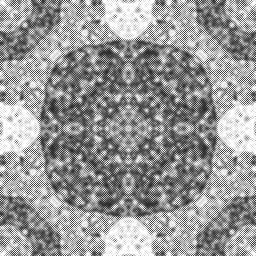}
\includegraphics[width=4.5cm,height=4.5cm]{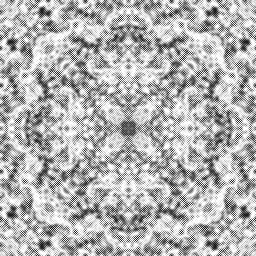}
\includegraphics[width=4.5cm,height=4.5cm]{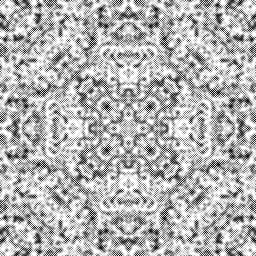}
\caption{Grayscale shaded contour graphics representing the values of the field $\psi$ after
3000 time steps for $d = 0.25$, periodic boundary conditions, three values of $c$ and type (B) initial conditions;
(a) $c = 3.8$, (b) $c = 3.9$, (c) $c = 4.0$.
Brighter regions are those with higher values of $\psi$.}
\end{figure}

\begin{figure}[h!]
\includegraphics[width=4.5cm,height=4.5cm]{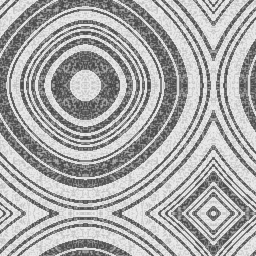}
\includegraphics[width=4.5cm,height=4.5cm]{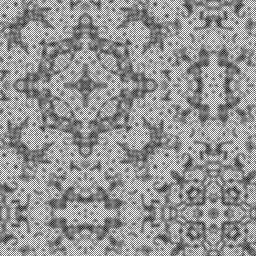}
\includegraphics[width=4.5cm,height=4.5cm]{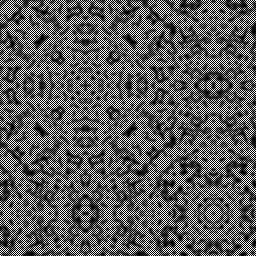}
\caption{Grayscale shaded contour graphics representing the values of the field $\psi$ after
3000 time steps for periodic boundary conditions and type (E) initial conditions;
(a) $c = 3.7$, $d = 0.05$, (b) $c = 3.9$, $d = 0.25$, (c) $c = 4.0$, $d = 0.25$.}
\end{figure}

In Figs. 16-22 we show shaded-contour plots representing the values of the field
$\psi(x,y)$ after 3000 time steps for Dirichlet boundary conditions and five types of initial conditions. 

\begin{figure}
\includegraphics[width=4.5cm,height=4.5cm]{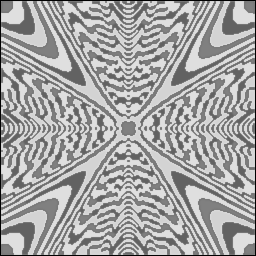}
\includegraphics[width=4.5cm,height=4.5cm]{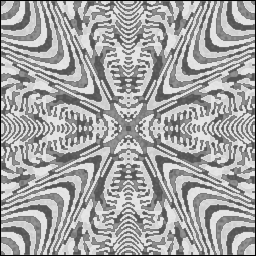}
\includegraphics[width=4.5cm,height=4.5cm]{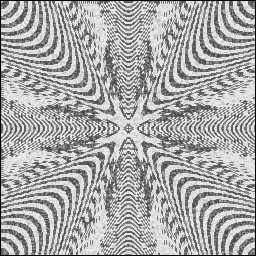}
\caption{Same as in Figure 9 but for Dirichlet boundary conditions.}
\end{figure}

Naturally, the large structures visible in the figures reflect, to some extent, 
the symmetry of the simulation box. More interesting observation is that the change 
from periodic ($c = 3.5$) to chaotic ($c = 3.6$) regime - 
as defined for individual maps -  does not lead,  
in the case of very slow diffusion, to any spectacular change of the pattern.

\begin{figure}[h!]
\includegraphics[width=4.5cm,height=4.5cm]{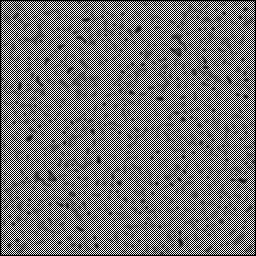}
\includegraphics[width=4.5cm,height=4.5cm]{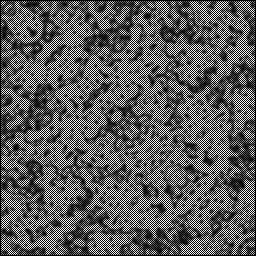}
\includegraphics[width=4.5cm,height=4.5cm]{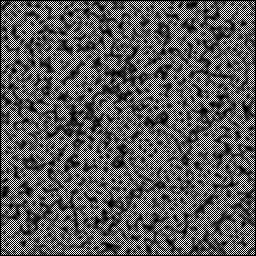}
\caption{Same as in Figure 10 but for Dirichlet boundary conditions.}
\end{figure}

\begin{figure}[h!]
\includegraphics[width=4.5cm,height=4.5cm]{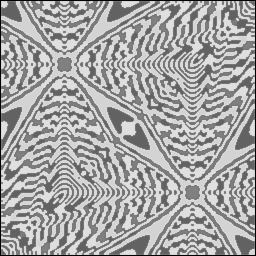}
\includegraphics[width=4.5cm,height=4.5cm]{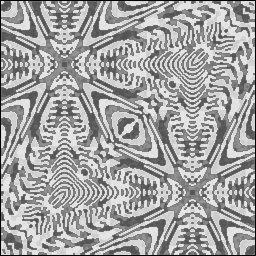}
\includegraphics[width=4.5cm,height=4.5cm]{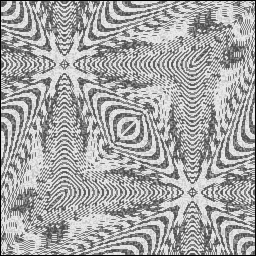}
\caption{Same as in Figure 11 but for Dirichlet boundary conditions.}
\end{figure}

\begin{figure}[h!]
\includegraphics[width=4.5cm,height=4.5cm]{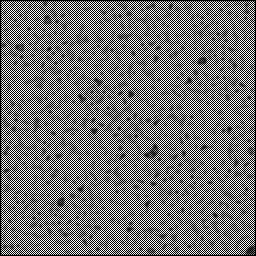}
\includegraphics[width=4.5cm,height=4.5cm]{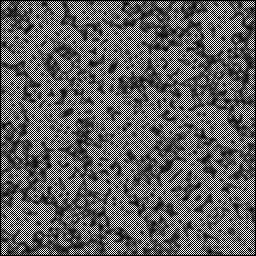}
\includegraphics[width=4.5cm,height=4.5cm]{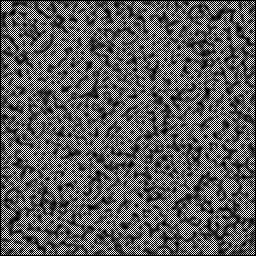}
\caption{Same as in Figure 12 but for Dirichlet boundary conditions.}
\end{figure}

\begin{figure}[h!]
\includegraphics[width=4.5cm,height=4.5cm]{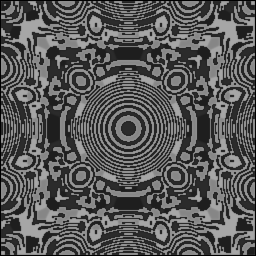}
\includegraphics[width=4.5cm,height=4.5cm]{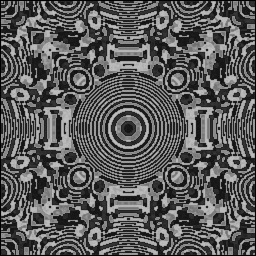}
\includegraphics[width=4.5cm,height=4.5cm]{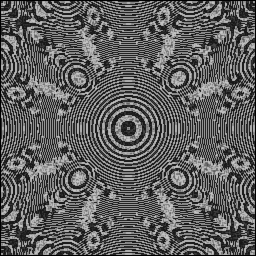}
\caption{Same as in Figure 13 but for Dirichlet boundary conditions.}
\end{figure}

The most characteristic feature of the fast-diffusion (i.e. large $d$) case is again 
the disappearance of the large-scale structures for all types of initial conditions. 
This is also reflected by flat curves in the plots of reduced wave functions.  
However, somewhat more pronounced grainy structures again reappear for $c = 3.9$
and large $d$.



\section{Concluding remarks}

Perhaps the most interesting of the various features of the considered system of coupled map lattices
is that it appears to be ``condensed'' if the most standard measures of the classical field theory of Bose condensates
are applied. That is, for a majority of parameter values we have observed that a gap between the largest
eigenvalue of the reduced density matrix and the rest has been developed.  
Only for $d = 0.25$ we have observed not a single,
but rather two eigenvalues which are much larger than all remaining ones. 
The latter fact might be of independent interest, as it seems to correspond with 
the so-called ``quasi-condensates".
Secondly, the prominent characteristic of the system is the presence of large-scale patterns for smaller
values of the ``diffusion constant" $d$, $d \leq 0.2$  and not too large values of the non-linear parameter,
$c \leq 3.8$. Thirdly, a very strong dependence of both the presence and qualitative features of the
patterns on the initial conditions is to be noticed. The latter fact should be a warning against restricting oneself
to one type of initial conditions - namely, the purely random ones - which is very often met in the literature.
The most interesting facts can be overlooked this way. Interestingly, the strong dependence of patterns
on the initial conditions takes place even in the non-chaotic regime of the parameter $c$. 
Fourthly, for very slow diffusion ($d = 0.05$) we have found that the ``number of particles'' -  
defined in a natural way - is an approximate constant of motion for sufficiently large times (because
the period-2 oscillations have very small amplitude). If the system exhibits period-2 or period-4 oscillations, 
the number of particle fluctuates around two (or four) values, as if there were two (four) different systems.

We have, in addition, performed similar numerical experiments with another version of the logistic map,
reaching similar conclusions. The same statement seems to be valid
in the case of standard (rather than logistic) map employed as a basis for the coupled map lattice; however,
we have only very preliminary results in that case.

Finally, we would like to observe that the domination of zeroth mode in the momentum space suggests that a kind
of Bogoliubov approximation could be applicable. This might lead to an efficient analytical approach
to the dynamics of CML.

We plan to develop further our attempt of using classical field-theoretical concepts in coupled map lattices.
Work is in progress of their using in the case of three-dimensional CMLs based on logistic maps as well as
other physically more appealing discrete systems.

\vspace*{1cm}

{\bf Acknowledgments}
It is a pleasure to thank Professor Mariusz Gajda and Dr. Emilia Witkowska for offering several
helpful discussions


\vspace*{1cm}

\begin{thebibliography}{00}

\bibitem{CF}
{\it Dynamics of Coupled  Map Lattices and Related Spatially Extended Systems},
edited by J.R. Chazottes and B. Fernandez (Springer, New York 2005)

\bibitem{Ilachinski}
A. Ilachinski, 
{\it Cellular Automata. A Discrete Universe}
(World Scientific, Singapore 2001)

\bibitem{Kaneko1}
K. Kaneko, 
Prog.Theor. Phys. {\bf 72}, 480 (1984) 

\bibitem{WK}
I. Waller and R. Kapral,
Phys. Rev. A {\bf 30}, 2047 (1984)

\bibitem{Kapral}
R. Kapral, 
Phys. Rev. A {\bf 31}, 3868 (1985)

\bibitem{Kaneko2}
K. Kaneko, 
Physica D {\bf 34}, 1 (1989)

\bibitem{YK1}
T. Yanagita and K. Kaneko,
Physica D {\bf 82}, 288 (1995)

\bibitem{Yanagita}
T. Yanagita,
Phys. Lett. A {\bf 165}, 405 (1992)

\bibitem{GC}
P.S. Ghoshdastidar and I. Chakraborty,
J. Heat Transfer {\bf 129}, 1737 (2007)

\bibitem{YK2}
T. Yanagita and K. Kaneko,
Phys. Rev. Lett. {\bf 78}, 4297 (1997)

\bibitem{Kaneko3}
K. Kaneko, in: {\it Pattern Dynamics, Information Flow, and Thermodynamics of Spatiotemporal Chaos},
edited by K. Kawasaki, A. Onuki, and M. Suzuki (World Scientific, Singapore 1990)

\bibitem{MFMF}
P. Muruganandam, F. Francisco, M. de Menezes, and F.F. Ferreira,
Chaos, Solitons and Fractals {\bf 41}, 997 (2009)

\bibitem{DGPS}
F. Dalfovo, S. Giorgini, L.P. Pitaevskii, and S. Stringari,
Rev. Mod. Phys. {\bf 71}, 463 (1999)

\bibitem{Leggett}
A. Leggett,
Rev. Mod. Phys. {\bf 73}, 307 (2001)

\bibitem{GGR1}
K. G\'oral, M. Gajda, and K. Rz\k{a}\.zewski, 
Opt. Express {\bf 8}, 92 (2001)

\bibitem{GGR2}
K. G\'oral, M. Gajda, and K.Rz\k{a}\.zewski,
Phys. Rev. A {\bf 66}, 051602(R) (2002)

\bibitem{GGR3}
K. G\'oral, M. Gajda, and K.Rz\k{a}\.zewski,
J. Opt. B: Quantum Semiclassical Opt. {\bf 5}, 96 (2003)

\bibitem{Sinha}
S. Sinha, 
Physical Review E {\bf 53} 4509 (1996) 

\bibitem{Janosi}
I.M. Janosi, L. Flepp, and T. Tel, 
Physial Review Letters {\bf 73}, 529 (1994)

\bibitem{BB}
G. Bianconi and A.-L. Barabasi,
Phys. Rev. Lett. {\bf 86}, 5632 (2001)

\bibitem{RB}
A. Reka and A.-L. Barabasi,
Rev. Mod. Phys. {\bf 74}, 47 (2002)

\bibitem{PO}
O. Penrose and L. Onsager, 
Phys. Rev. {\bf 104}, 576 (1956)

\bibitem{Yang}
C.N. Yang, Rev. Mod. Phys. 
{\bf 34}, 694 (1962)

\bibitem{KGR}
D. Kadio, M. Gajda, and K. Rz\k{a}\.zewski,
Phys. Rev. A {\bf 72}, 013607 (2005)

\bibitem{PSW1}
D.S. Petrov, G.V. Shlyapnikov, and J.T.M. Walraven,
Phys. Rev. Lett. {\bf 85}, 3745 (2000)

\bibitem{PSW2}
D.S. Petrov, G.V. Shlyapnikov, and J.T.M. Walraven,
Phys. Rev. Lett. {\bf 87}, 050404 (2001)



\end{thebibliography}
\end{document}